\documentclass{article}
\usepackage{graphicx}

\begin{document}
\title{Role of Site-selective Doping on Melting Point of CuTi Alloys: A Classical Molecular Dynamics Simulation Study}

\author{Karabi Ghosh, \\Theoretical Physics Division, \\Bhabha Atomic Research Centre, Mumbai-400085, India\\ \\
 Manoranjan Ghosh, \\Technical Physics Division, \\Bhabha Atomic Research Centre, Mumbai-400085, India\\ \\
 S.V.G. Menon,\\Theoretical Physics Division, \\Bhabha Atomic Research Centre, Mumbai-400085, India}
\date{}
\maketitle
\begin{abstract}
Effect of site-selective substitution of Ti in Cu on the thermal stability of CuTi alloy is investigated using classical molecular dynamics simulations with Embedded Atom Method potentials. It has been observed experimentally that melting point of all the naturally occurring stable phases of CuTi alloys do not show a definite trend with gradual increase in Ti concentration. To understand the phenomenon, super cells of CuTi alloy are constructed where Cu atom is substituted by Ti randomly and at selective sites.  For random substitution, the melting point decreases linearly with increase in Ti concentration. A non-monotonous dependence is seen when Cu atoms at selective sites are replaced by Ti. For a particular doping concentration, the melting point shows a  wide range of variation depending on the order of atomic arrangement, and can be fine tuned by selecting the sites for substitution. The variations in melting points in different cases are explained in terms of the peak height, width and position of the corresponding radial distribution functions. Finally, it is verified that initial structures of the naturally occurring CuTi alloys are responsible for the non-definite trend in their melting points.  

\end{abstract}



\section{Introduction}
Identification of doping strategies is important to optimize the physical properties of alloys . Usually dopant concentration is varied to tailor the properties. For a fixed concentration, position of the dopant atom in the host matrix is also a key factor as it determines the interaction between host and dopant atoms. Thus doping at the selective sites is a viable tool to achieve finer control over the properties of the alloy. Site selective doping has been realized experimentally and is also used to make an educated guess of the material properties. For example, spatially controlled doping of a set of impurity atoms (cobalt, silver, nitrogen, and boron) in carbon nanotube has been achieved experimentally by ion implantation technique \cite{Bangert}. $ \mathrm{Lu_2SiO_5}$ has been doped by $\mathrm{Eu^{3+}}$ at preferential site by sol-gel chemistry method \cite{Mansuy}. Site selective substitution of boron in carbon nanoribbons has been investigated in the framework of first-principles density functional theory. Electronic properties are obtained for different boron$-$boron arrangements and concentrations \cite{Navarro-Santos}. While majority of these reports are focused on the electronic properties, role of site selective doping on the thermal stability of the alloy is rarely investigated. 

In the present work we investigate the role of site selective substitution of Ti in Cu on the melting point of CuTi alloy. CuTi is technologically significant due to its high strength and conductivity. It has high yield strength when hardened via precipitation, is widely used for direct brazing and joining of ceramics \cite{Li} and considered as a promising material for thermoelectric power generation \cite{Rathnayaka}. There are three main aspects of studying variation in melting point of CuTi alloys. 

The cell volume of some alloys increases compared to that in its pure phase due to the larger radius of the dopant atoms. Melting point of these alloys decrease if the melt occupies larger volume than its solid phase. CuTi alloy belongs to this category and the melting point of CuTi alloy should decrease as Ti concentration increases. But the phase diagram of CuTi alloy shows decrease in melting point, from 1356 K to 1158 K, on alloying by Ti only below a certain concentration (20 \% in $\mathrm{Cu_4Ti}$) \cite{Soffa}. Beyond this limit, stable phases are found at much elevated temperatures. Melting point gradually increases from 1158 K to 1273 K as the Ti concentration increases for stable phases like $\mathrm{Cu_2Ti}$, $\mathrm{Cu_3Ti_2}$, $\mathrm{Cu_4Ti_3}$ and CuTi (50 \%). Therefore linear dependence of melting point on Ti concentration is not observed. This anomaly in the melting curve can be explained by performing simulations on thermal stability of CuTi alloy having various arrangements of dopant atoms.

Crystals of CuTi alloys can be prepared over a wide range of Ti concentration by heat treatment and quenching of the melt of pure Cu and Ti. Cu-Ti phases having different thermal stabilities can be obtained by changing the temperature and duration of heat treatment \cite{Yoshimoto},\cite{LiLi}. Also, hardening of CuTi alloys with low Ti concentration, via aging is an important technique to enhance its ductility. Both the methods of preparation and processing, like age hardening of different phases of CuTi alloy, require knowledge of melting point in advance.

There are issues related to precipitation during age hardening of alloys to create high yield strength. In this process clusters of dopant atoms are created by cooling the heat treated alloy for forming mobile defects. Since these kinds of defects have large impact on the melting point, their creation requires prior information on the thermal stability of the alloy.

The organization of the paper is as follows: In section \ref{details}, we discuss classical molecular dynamics method for studying the CuTi system. The one phase method for determining the melting point, the embedded atom method (EAM) potential and the different initial structures obtained through doping are introduced. In the section \ref{results} we discuss the impact of random substitution, impurity atom cluster/micro-structure formation and site-selective substitution on the melting point of the alloy. The anomaly in the naturally existing stable phases of CuTi alloys is also considered. The variations in melting point are correlated with the changes in peak position, height and width of the radial distribution function (RDF) . Finally our results are summarized in section \ref{conclu}. 
\label{intro}

\section{Computational Details}
\label{details}
\subsection{Classical Molecular Dynamics simulation}
\label{MD}
The melting points of the CuTi alloy are determined using classical molecular dynamics simulations \cite{MDsimulation}. We employ the parallel molecular dynamics simulation package $\mathrm{DL_-POLY}$ \cite{DLPOLY}. $\mathrm{DL_-POLY}$ has been used earlier to model the melting process of the inter-metallic compound $\mathrm{Ni_3Al}$ \cite{NiAl}. We consider cubic super cells consisting of $11\times 11\times 11$ conventional FCC unit cells, which corresponds to $5324$ atoms. Three dimensional parallelopiped periodic boundary conditions are applied on the super cells to eliminate the surface effects and reproduce the bulk properties. The Berendsen isothermal-isobaric (NPT) ensemble is used to achieve constant temperature and pressure conditions \cite{berend}. The relaxation times for the thermostat and barostat are $1.0$ and $3.0$ ps, respectively, and the pressure is fixed at 0 atm. We use the Verlet leapfrog scheme for integrating the Newton's equations of motion \cite{Verlet} with a time step of 0.001 ps for all cases. Simulations are done for a total of 10,000 time steps, where the first 4000 steps are used for equilibration and the remaining 6000 for statistical averaging.  In order to ascertain that only one image of a particle interacts with another particle, the cut off distances for force calculations is chosen to be smaller than half the size of the supercell. A cutoff distance of 10\r{A} is used for all the simulations. 
\subsection{Determination of melting point using one phase method}
\label{onephase}
\begin{figure}
\includegraphics[width=9cm]{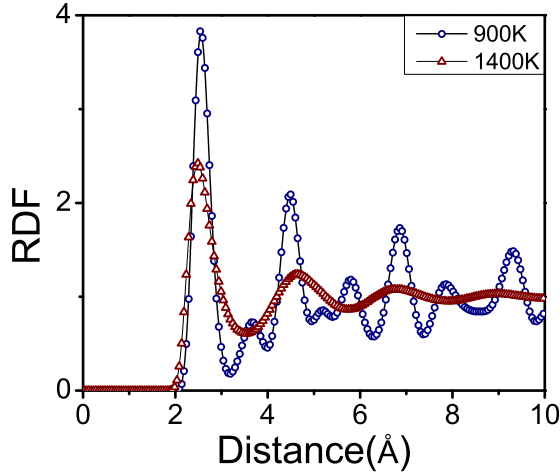}
\caption{\label{rdf}Radial distribution function (RDF) of Cu before and after melting. }
\end{figure}

\begin{figure}
\includegraphics[width=9cm]{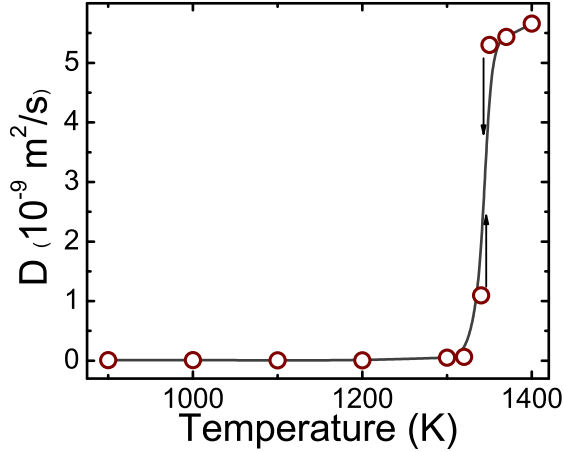}
\caption{\label{diff-coeff}Jump in the diffusion coefficient (D) of Cu at melting point. }
\end{figure}

The melting point of the CuTi systems is determined using the one-phase method \cite{onephase} wherein the melting point is identified from sharp increase in atomic volume, diffusion coefficient and energy as temperature is varied. Initially the temperature is increased by 100K  until the alloy melts and the range within which the jumps occur is observed. The structure obtained after equilibration at a certain temperature is used as the starting configuration for the subsequent higher temperature. The melting point for the specific alloy is next obtained by increasing the temperatures by 10K and observing the jumps in the above mentioned quantities. Also, as the long range order is lost once melting occurs, the height of the first peak of RDF decreases drastically and RDF $\approx$ 1 after 2-3 small humps. The RDF describes how the atomic density varies as a function of distance from the atom and is given by $g(r)=n(r)/(4 \pi r^2 \Delta r \rho)$ where $n(r)$ is the number of atoms at a distance $r$ within a shell of thickness $\Delta r$ and $\rho$ is the average number density. The heights of the peaks signify the number of first, second, etc., nearest neighbours. Similarly the position of RDF peaks reflects the neighbour distances. Earlier, a similar study of alloying bcc Fe with Si has been performed and the increase in melting point is explained in terms of the RDF peak height \cite{Belonoshko}.  For Cu, at a temperature of 1440K, the intermediate and long distance peaks of Cu-Cu RDF are seen to slowly merge out indicating that melting has occured (Fig. \ref{rdf}). The jump in diffusion coefficient for Cu is shown in Fig. \ref{diff-coeff} wherein the melting point 1340K is indicated using arrows. For all the doped CuTi alloys, the melting points are obtained in a similar manner.

\subsection{Embedded Atom Method Potential}
\label{EAM}
We have accounted for the interaction between atoms using the Embedded Atom Method (EAM) potential which incorporates the many-atom interactions neglected within the pair potential scheme. The EAM provides a semi-empirical potential for computing the total energy of a metallic system \cite{Daw}. In this scheme, we view the energy of the metal as obtained by embedding an atom into the local electron density provided by the remaining atoms, in addition to the pair interaction. It is especially useful for systems with large unit cells and is appropriate for metals with empty or filled d-bands \cite{Dawrev}. Thus the total internal energy, which consists of an embedding term plus the usual two body interactions, is expressed as \cite{Hong}
\begin{eqnarray}
E_{tot}=\sum_{i} F(\rho_i)+\frac{1}{2}\sum_{i}\sum_{j \neq i}\phi_{ij}(r_{ij}),\\
\rho_i=\sum_{j \neq i}f(r_{ij}),
\end{eqnarray}
where $\rho_i$ is the local electron density at  $i^{th}$ atom due to all the other atoms, $\phi(r_{ij})$ is the pair potential and $r_{ij}$ is the distance between $i^{th}$ and $j^{th}$ atoms. The embedding energy, $F(\rho_i)$, which is a function of the local-electron density, is approximated by superposing the atomic electron density functions, $f(r_{ij})$. The two-body potential and electron density functions, as given by Johnson, are respectively written as
\begin{eqnarray}
\phi(r)=\frac{Ae^{-\alpha(\frac{r}{r_e}-1)}}{1+(\frac{r}{r_e}-\kappa)^m}-\frac{Be^{-\beta(\frac{r}{r_e}-1)}}{1+(\frac{r}{r_e}-\lambda)^n}\\
f(r)=\frac{f_ee^{-\beta (\frac{r}{r_e}-1)}}{1+(\frac{r}{r_e}-\lambda)^n}
\end{eqnarray}
The pair term is a Morse-type potential which is composed of a short-range repulsive exponential and a long-range attractive exponential. The parameters in Eqs. (3)-(4) are given in Table \ref{table1}. The embedding function, F, is given by
\begin{eqnarray}
F(\rho)=F_e(1-ln(\frac{\rho}{\rho_e})^\eta)(\frac{\rho}{\rho_e})^\eta
\end{eqnarray}
where $F_e$ is the embedding function corresponding to equilibrium electron density $\rho_e$. A smoother embedding function that matches physical properties better than Eq. (5) is given by \cite{Wadley}
\begin{eqnarray}
F(\rho)&=& \sum_{i=0}^{3}F_{ni}(\rho/\rho_n-1)^i,\ \rho<\rho_n=0.85\rho_e, \\
F(\rho)&=& \sum_{i=0}^{3}F_{i}(\rho/\rho_e-1)^i,\ \rho_n\leq \rho<\rho_0=1.15\rho_e, \\
F(\rho)&=& F_0(1-ln(\frac{\rho}{\rho_e})^\eta)(\frac{\rho}{\rho_e})^\eta,\ \rho_0<\rho
\end{eqnarray}
The parameters in Eqs. (6)-(8) are given in Table \ref{table2}. 

For an alloy model using EAM, an embedding function $F(\rho)$ and an atomic electron-density function $f(r)$ are to be specified for each atomic species, while two-body potentials are needed for each possible pairs of atomic species. Since the electron density at any location is taken as a linear superposition of atomic electron densities, and the embedding energy is assumed to be independent of the source of the electron density, these two functions can be taken from mono atomic models directly. For a binary alloy with a- and b-type atoms, $\phi^{aa}$ and $\phi^{bb}$ are given by the individual mono atomic models, and $\phi^{ab}$ and $\phi^{ba}$ are assumed to be equal. A direct and simple possible form for the alloy potential can assume either the geometric or arithmetic average of two-body potentials as has been shown by Foiles et al. \cite{Foiles}. However, a novel two-body alloy model, which satisfies the invariance properties of electron density in a fcc mono atomic metal, has been constructed by Johnson \cite{Johnson}. In this work, we adopt Johnson's alloy model using EAM which satisfies the invariance condition:
\begin{eqnarray}
\phi^{ab}(r)=\frac{1}{2}[\frac{f^b(r)}{f^a(r)}\phi^{aa}(r)+\frac{f^a(r)}{f^b(r)}\phi^{bb}(r)]
\end{eqnarray}
where $\phi^{ab}$ is the alloy potential, f is the density function and the superscripts aa and bb stand for mono atomic a and b, respectively.

\begin{table}
\caption{\label{table1} Parameters of electron density function and two-body potential for Cu and Ti.}

\begin{tabular}{ccc}
\hline
& Cu & Ti\\
\hline 
$r_e$(\r{A}) & 2.556 & 2.930\\
$f_e$(eV/\r{A}) & 1.554 & 1.860\\

$\alpha$  & 7.670 & 8.780\\
$\beta$ & 4.09 & 4.68\\
A & 0.328 & 0.328\\
B & 0.469 & 0.469\\
$\kappa$ & 0.431 & 0.431\\
$\lambda$ & 0.863 & 0.863\\ 
m & 20 & 20\\
n & 20 & 20\\
\hline
\end{tabular}
\end{table}

\begin{table}
\caption{\label{table2} Parameters of embedding function for Cu and Ti.}
\begin{tabular}{ccc}
\hline
& Cu & Ti\\
\hline 
$\eta$ & 0.921 & 0.560\\
$\rho_e$(eV/\r{A}) & 22.150 & 25.600\\
$F_{n0}$(eV) & -2.176 & -3.200\\
$F_{n1}$(eV) & -0.140 & -0.200\\
$F_{n2}$(eV)  & 0.286 & 0.680\\
$F_{n3}$(eV) & -1.751 & -2.320\\
$F_0$(eV)& -2.190 & -3.220\\
$F_1$(eV) & 0.000 & 0.000\\
$F_2$(eV) & 0.703 & 0.610\\
$F_3$(eV) & 0.684 & -0.750\\ 
\hline
\end{tabular}
\end{table}

\subsection{Initial CuTi alloy structure}\label{initial}
Structures of CuTi alloys having different Ti concentrations have been generated by random or selective substitution of Ti in perfect fcc Cu supercells. The origin of the cartesian coordinate system is taken to be the centre of the supercell. The perfect fcc Cu supercell is generated using the program genlat.f in utility of $\mathrm{DL_-POLY}$. The Cu atoms are placed one after the other starting from the (-,-,-) octant towards the (+,+,+) octant.

For random doping of Cu atoms with Ti, random numbers (depending on the concentration) lying within the supercells are generated for x, y and z coordinates. The Cu atom whose coordinates are nearest to the random numbers are replaced by Ti atoms.  

 Microstructures of Ti having various sizes are also generated within the Cu supercells. Single microstructures of different concentrations are generated by replacing all the Cu atoms with Ti atoms inside spheres of varying radii at the centre of the microstructures i.e., (0,0,0). For substituting 8 microstructures, all Cu atoms lying within spheres of different radii centered at (-D,-D,-D), (-D,-D,+D), (-D,+D,-D), (+D,-D,-D), (-D,+D,+D), (+D,-D,+D), (+D,+D,-D) and (+D,+D,+D) with D=10.845\r{A} are replaced by Ti atoms. Similarly, 9 microstructures are obtained by replacing the Cu atoms within the sphere at the origin in addition to the spheres used for 8 microstructures.
 
Selective doping is done in a variety of ways: For a concentration of 5 \% Ti, first Cu atom among every 20 atoms is replaced by Ti till all the 5324 atoms are covered. The selectively doped CuTi alloy generated in this way is called selective 5 \% atom1. Similarly for a concentration of 10 \% Ti, first Cu atom among every 10 atoms is replaced by Ti. This alloy is named as selective 10 \% atom1. A type named atom2 is generated by replacing 2 atoms at a time. So, for a concentration of 5 \% Ti, first two Cu atoms among every 40 atoms are replaced by Ti. Thus, selective 33.33 \% atom1 is generated by replacing first among every 3 Cu atoms with Ti and selective 33.33 \% atom6 by replacing first six among every 18 Cu atoms with Ti.  

The natural CuTi alloy structures, viz. CuTi and $\mathrm{CuTi_2}$ are also generated using the program genlat.f. The number of atoms of Cu and Ti and their positions within the unit cells are obtained from ICSD database \cite{ICSD}.

\section{Results and Discussions}\label{results}
\subsection{Random doping}
\begin{figure}
\begin{center}
\includegraphics[width=9cm]{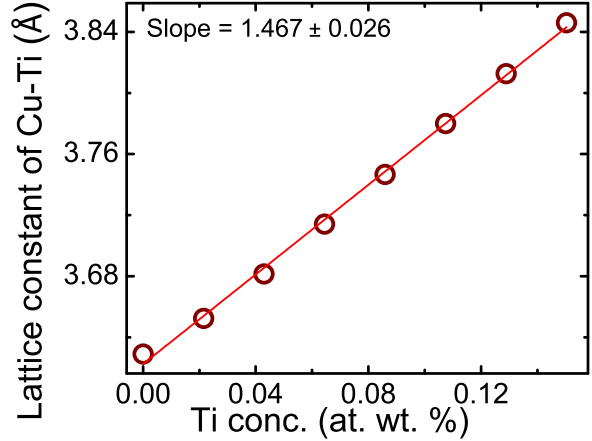}
\end{center}
\caption{\label{vegards} Linear variation of the lattice parameter as a function of the atomic weight percent of the dopant Ti in Cu.}
\end{figure}

\begin{figure}
\begin{center}
\includegraphics[width=9cm]{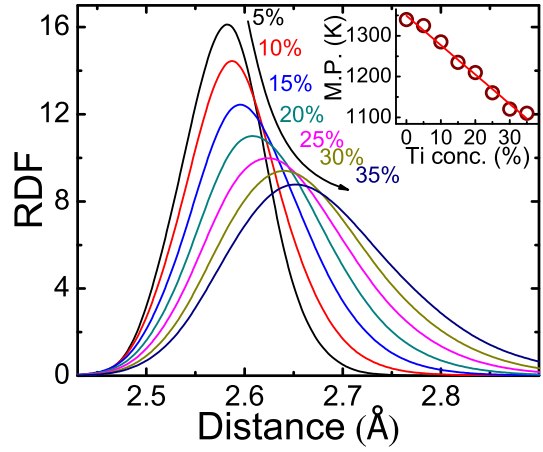}
\end{center}
\caption{\label{rand-rdf} RDFs for Cu-Cu bond in the case of random doping of Cu with Ti. Inset shows linear variation in melting point (M.P.) as a function of the number percent of the dopant Ti in Cu.}
\end{figure}

The melting temperature for Cu is reported as 1200K in an earlier study employing $\mathrm{DL_-POLY}$ with Sutton Chen potential \cite{onephase}. However, we get 1340 K which agrees better with the experimental result of 1356K. For random doping, melting point decreases linearly as Ti concentration increases (inset of Fig. \ref{rand-rdf}). Since the atomic radius of Ti (2 $\mathrm{\dot A}$) is higher than that of Cu (1.57 $\mathrm{\dot A}$), cell volume increases when a Ti atom replaces a Cu atom. The increase in lattice parameter of the CuTi alloy is found to obey the empirical Vegard's law, which, for a given temperature, is a linear relation between lattice constant and concentration of the constituent elements \cite{vegards} (see Fig. \ref{vegards} ). As a result, the average distance between Cu atoms increases and the Cu-Cu bond strength decreases. This is clearly reflected in Fig. \ref{rand-rdf} which shows the 0K RDF for the Cu-Cu bond. It is observed that the heights of the RDF peaks decrease, full widths at half maxima (FWHM) increase and the peak positions shift to the right. As mentioned earlier, decrease in RDF peak height is directly linked to the reduction in number of nearest neighbours.  Also, increase in cell volume leads to shifting of RDF peak to the right. Thus for random doping, the average Cu-Cu bond strength and hence melting point decreases linearly as dopant concentration increases.

\subsection{Microstructural doping}

\begin{figure}
\begin{center}
\includegraphics[width=9cm]{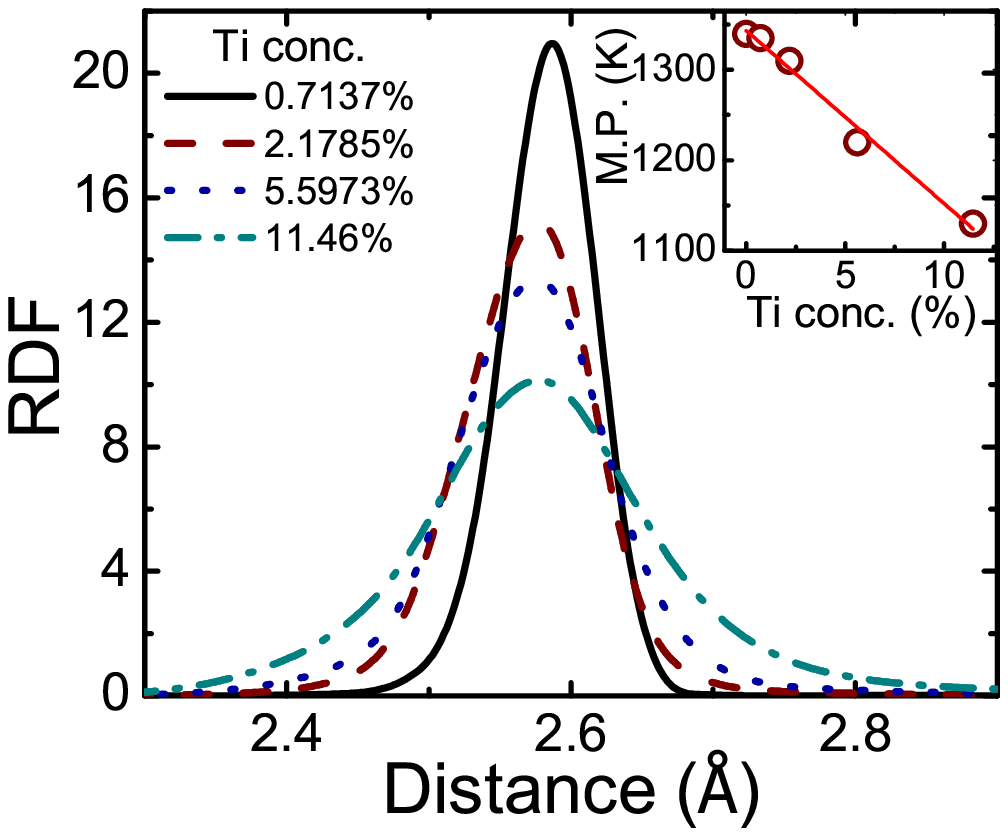}
\end{center}
\caption{\label{micro-rdf} RDFs for Cu-Cu bond in case of single microstructure doping of Cu with Ti. Inset shows linear variation in melting point as a function of the number percent of the dopant Ti in Cu.}
\end{figure}  

\begin{figure}
\begin{center}
\includegraphics[width=9cm]{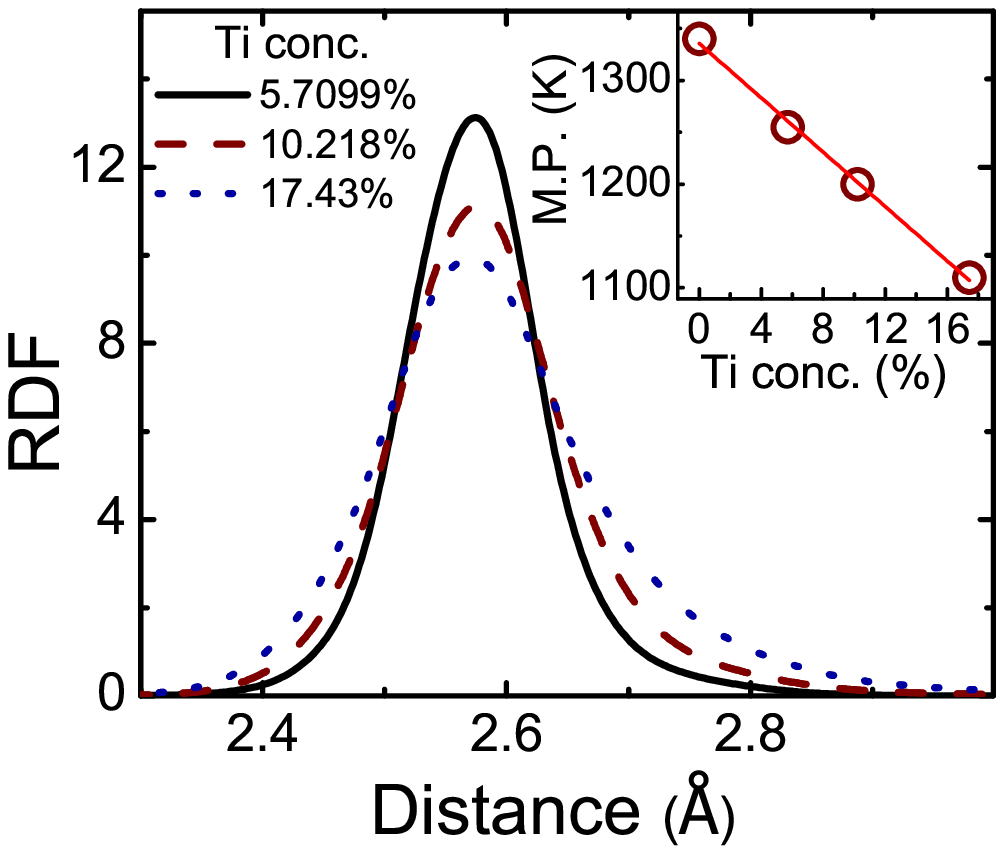}
\end{center}
\caption{\label{8micro-rdf} RDFs for Cu-Cu bond in case of 8 microstructure doping of Cu with Ti. Inset shows linear variation in melting point as a function of the number percent of the dopant Ti in Cu.}
\end{figure}  

For most of the practical cases, doping of clusters of several atoms is more probable compared to atom by atom substitution. Therefore, micro-structures of several Ti atoms are created within the Cu lattice . For a single microstructure, the Cu-Cu bond strength decreases linearly with increase in Ti concentration as shown in Fig.\ref{micro-rdf}. This is also reflected in the linear decrease of melting point (inset of Fig. \ref{micro-rdf}). Similarly, for 8 microstructure doping, the RDF peak height decreases linearly (Fig. \ref{8micro-rdf}) and a gradual decrease in melting point is observed (see inset of Fig. \ref{8micro-rdf}).  Exactly similar trends are observed for 9 microstructure doping.

\subsection{Selective doping}
\begin{figure}
\begin{center}
\includegraphics[width=9cm]{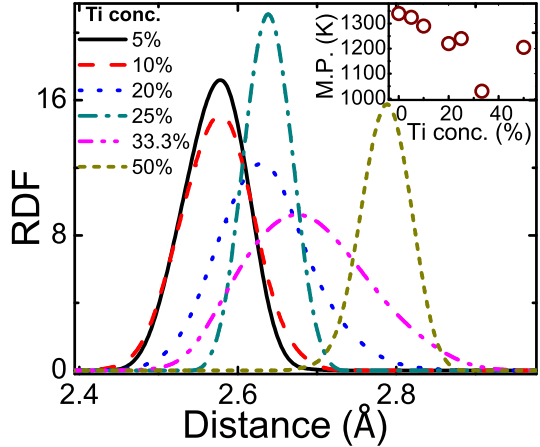}
\end{center}
\caption{\label{selective}RDFs for Cu-Cu bond in case of selective doping of Cu with Ti (atom1). Inset shows variation in melting points as a function of the number percent of the dopant Ti in Cu.}
\end{figure}

The systematic variation in melting point on changing the Ti concentration as observed for random and microstructural doping is no longer seen in case of selective dopant substitution. Both dopant concentration and the site of substitution are responsible for determining the melting point of the alloy. As already pointed out, Ti atoms can be introduced into Cu lattice in various ways to generate different atomic arrangements having the same dopant concentration. When a Ti atom replaces a single Cu atom in a unit (atom1), melting point decreases with Ti concentration up to 20 \% as in the case of random substitution (inset of Fig. \ref{selective}). Then the melting point increases for 25\% Ti. As shown in Fig. \ref{selective}, upto 20 \% Ti concentration, the height of the first RDF peak decreases, become broader and position shifts to the right showing a loss in symmetry of the structure. But for 25\% of Ti concentration, the first RDF peak becomes narrower and attains its maximum value. Its position does not shift further indicating stronger Cu-Cu bonding for the structure as compared to 20 \% Ti. Therefore, high melting point observed in case of 25 \% atom1 case can be understood in terms of the height, width and position of the RDF peak. In a similar manner, the low melting point found in case of 33.33 \% atom1 case can be understood by the loss in Cu-Cu bond strength which is clearly evident from the corresponding short and broad RDF peak. Finally, peak height increases and becomes narrower for 50 \% Ti which results in increase in melting point. Similar kind of non-monotonous behaviour of melting point with dopant concentration can be seen for selective atom2 to atom6 cases. Only the late rise in melting point can be seen for different dopant concentration depending on the initial atomic arrangement.

The correlation between the melting point and characteristics of the RDF peak established here is important for the following reason. For an alloy it is necessary to quantify the interaction between the host atoms as well as between the host and dopant atoms. Especially, at higher temperatures when atoms may be displaced from their equilibrium positions, characteristics of the RDF peaks can predict the symmetry of the atomic configuration and the thermal behaviour of the alloy.

\begin{figure}
\begin{center}
\includegraphics[width=9cm]{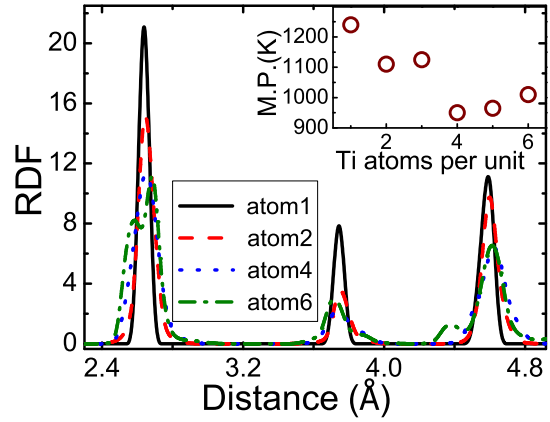}
\end{center}
\caption{\label{select-25-rdf}First three peaks of the Cu-Cu RDF for selective substitutional doping of Cu with 25\% Ti doping. Inset shows variation in melting points for different Ti arrangements having 25\% Ti concentration.}
\end{figure}
 
The role of site selective doping on thermal stability can be unequivocally established if melting point is shown to change for different arrangement of dopant atoms but having same concentration. For that purpose, six different configurations of 25\% doped CuTi alloy having different arrangements of Ti atoms in Cu lattice are constructed. Variation of the melting point for different Ti arrangements is depicted in the inset of Fig. \ref{select-25-rdf}. The first, second and third Cu-Cu RDF peaks for four arrangements, namely, atom1, atom2, atom4 and atom6 are shown in Fig. \ref{select-25-rdf}. The melting point is seen to decrease upto atom4 and then increases for atom5 and atom6. As shown in Fig. \ref{select-25-rdf}, up to atom4, the height of the first RDF peak decreases showing a loss in symmetry of the structure. However, for atom6, the RDF peak height increases and shifts to the left showing a more compact structure thus explaining the increase in melting point.

\begin{figure}
\begin{center}
\includegraphics[width=9cm]{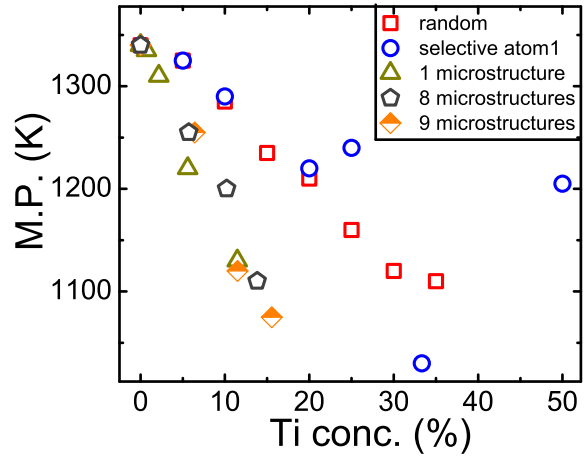}
\end{center}
\caption{\label{mp}Melting points obtained for different types of substitutional doping of Cu with Ti.}
\end{figure}

Finally, we plot the melting points against dopant concentration for different ways of doping, namely, random, microstructural and selective doping (Fig. \ref{mp}). For random, single, 8 and 9 microstructures, linear variation of melting point on Ti concentration is observed. However, in case of microstructural doping, the melting points decrease faster compared to random doping. Selective doping (atom1), on the other hand, does not show the linear variation and the melting point depends on the initial structure irrespective of the concentration. 
 
\subsection{Natural CuTi alloys}

\begin{figure}
\begin{center}
\includegraphics[width=9cm]{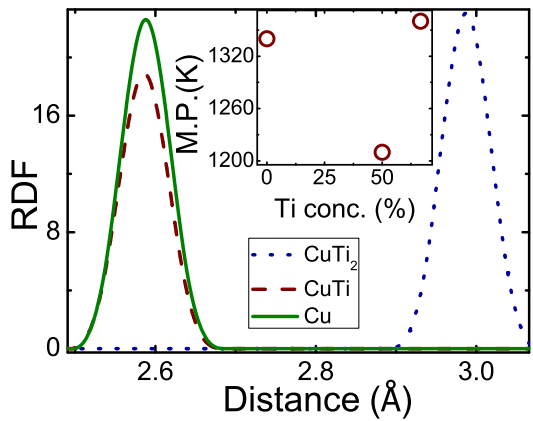}
\end{center}
\caption{\label{natural}First Cu-Cu RDF peaks for natural phases of CuTi alloy. Inset shows variation in melting points.}
\end{figure} 
The link established so far, between melting points and characteristics of the RDF peaks of an alloy, can be employed to understand the observed anomaly in the melting curve of naturally existing phases of CuTi alloys. Melting point as extracted from the phase diagram of CuTi alloy show non-monotonous dependence on Ti concentration. In the inset of Fig. \ref{natural}, the calculated melting points of Cu and two of its naturally occurring alloys viz. CuTi and $\mathrm{CuTi_2}$ are plotted and the corresponding RDF peaks are shown in Fig. \ref{natural}. In line with the experimental phase diagram, melting point Vs. Ti concentration first decreases and then again increases. For example, substantial reduction in the melting point of naturally occuring CuTi alloy (1210 K) is seen compared to pure Cu (1340 K). Melting point of $\mathrm{CuTi_2}$ increases to 1360K for 66.6 \% doping (see inset of Fig. \ref{natural}). The late rise in melting point with the increase of Ti concentration arises due to higher ordering between the atoms in $\mathrm{CuTi_2}$ (I4/mmm) than CuTi (P4/nmmS) . This is reflected on the Cu-Cu bond strength which can be visualized by the reduced RDF peak height of CuTi compared to pure Cu. As expected, RDF peak height of $\mathrm{CuTi_2}$ increases which explains its observed higher melting point.

\section{Conclusions}\label{conclu}

In summary, the role of site-selective substitution of Ti in Cu on the melting point of CuTi alloy has been investigated. Super cells of CuTi alloy having different arrangement of Ti atoms are constructed. Results obtained by replacing Cu atoms by Ti randomly, selectively and in the form of clusters are analyzed. We have established that, in addition to the concentration, the arrangement of dopant atoms in the host lattice plays a pivotal role in determining the melting point. A direct link between the melting point and characteristics of the RDF peaks of the alloy has been established. This facilitates to explain the variation in thermal stability in terms of the bond strength between host as well as host and dopant atoms in the alloy. The proposition has been validated by explaining the anomaly in the melting curve seen in naturally occurring phases of CuTi alloys having different crystal structures. The present study can be extended to other alloys of its kind and is useful for predicting doping strategies for fabrication of the alloy.

\end{document}